\documentstyle[twocolumn,psfig,aps]{revtex}
\begin{document}
\draft
\twocolumn[\hsize\textwidth\columnwidth\hsize\csname @twocolumnfalse\endcsname
%
%
%

\title{ Indications of Spin-Charge Separation in the 2D Extended t-J Model}

\author{George B. Martins$^1$, Robert Eder$^2$, and Elbio Dagotto$^1$}

\address{$^1$National High Magnetic Field Lab and Department of Physics,
Florida State University, Tallahassee, FL 32306}
\address{$^2$Institut f\"ur Theoretische Physik, Universit\"at W\"urzburg,
Am Hubland, 97074 W\"urzburg, Germany}

\date{\today}
\maketitle

\begin{abstract}
The 2D extended $\rm t-J$ model 
 is studied computationally in a broad region of parameter space,
motivated by recent 
photoemission experiments for the undoped cuprate $\rm Ca_2 Cu O_2 Cl_2$
(F. Ronning et al., Science {\bf 282}, 2067 (1998)).
The one-hole ground state is shown to develop
robust antiferromagnetic (AF) correlations between spins separated by
the mobile hole (i.e. across the hole). This effect tends to
decouple charge from spin, 
and the quasiparticle weight becomes negligible, particularly
at momenta $(0,\pi)-(\pi,0)$.
Studies with more holes show precursors of
metallic stripe formation, with holes sharing
their individual spin arrangements, and AF-correlations
generated across the stripe. 

\end{abstract}

\vskip2pc]
\narrowtext

Electronic strong
correlations are widely believed to be crucial for the explanation of the 
anomalous properties of high-temperature superconductors.
Several aspects of the cuprate phenomenology are indeed contained
in the two-dimensional (2D) $\rm t-J$ model, including 
tendencies towards
d-wave pairing upon hole-doping. 
However, pioneering angle-resolved photoemission (ARPES) studies
by Wells et al.\cite{wells} on the insulating compound
$\rm Sr_2 Cu O_2 Cl_2$ revealed important
discrepancies between
experimental data and $\rm t-J$ model predictions\cite{review}
near momenta
 ${\bf k} = (\pi,0)-(0,\pi)$, of relevance for the doped cuprates.
Moreover, recent ARPES experiments by Ronning et al.\cite{ronning}
for $\rm Ca_2 Cu O_2 Cl_2$ reported indications of a hole
dispersion with $\rm d_{x^2 - y^2}$-characteristics even
in such an undoped
compound, a remarkable and unexpected result. 
Clearly the one-hole case must be better understood
 before addressing a finite hole density.

In addition,
2D $\rm t-J$ model studies have not been able to stabilize the
metallic stripes proposed as an explanation of neutron
scattering experiments, with individual stripes affecting
only one $\rm CuO$ chain with hole 
density n=0.5\cite{tranquada}. Instead, a pattern
with stripes involving 
two adjacent chains has emerged using the DMRG method
with suitable boundary
conditions\cite{white}.
Since the study of stripes in the 2D $\rm t-J$ model seems a subtle
problem, 
it is important to develop new scenarios to generate
metallic stripes 
upon doping to guide the interpretation of experimental results.
With a similar motivation, Laughlin\cite{laughlin} recently
argued for the possible existence of a new, but still
unknown, fixed-point
in an extended parameter space that could influence
on the behavior of holes in antiferromagnets. Observing 
a new
fixed-point is potentially important for a proper  analysis of
the cuprates.

To search for theories beyond those currently available, a
 systematic analysis of the ``extended''
$\rm t-J$ model should be carried out. In this model, 
next-nearest-neighbor (NNN) hopping terms at 
distances 
$ \sqrt{2}a$ ($\rm t'$) and $2a$ ($\rm t''$) 
are added to the standard $\rm t-J$ model, which
only contains a nearest-neighbor amplitude t ($a$ is the  lattice 
spacing).
The importance of NNN-hoppings to reproduce ARPES results
was discussed by Nazarenko et al.\cite{nazarenko}, Belinicher et
al.\cite{bel}, Eder et
al.\cite{eder}, Kim et al.\cite{kim}, and 
addressed by other groups\cite{refer}. 
Tohyama et al. also remarked the importance of 
NNN-amplitudes\cite{tohyama}. The optimal values of $\rm t'/t,t''/t$
(-0.35 and 0.25, respectively\cite{eder}) are compatible with 
band structure calculations\cite{andersen}.
It is currently accepted 
that the extended $\rm t-J$ model
produces a quasiparticle
dispersion in excellent agreement with ARPES data.
However, intuition
is still lacking on the effect of
$\rm t',t''$ on the behavior of holes.

The main
motivation of this paper is to contribute to the clarification of the
physics contained in
 the 2D extended $\rm t-J$ model at low hole-density. 
Pursuing such goal, several surprises have
been found. The most interesting is the stabilization by
NNN-hoppings of a dynamically generated complex
structure around mobile holes containing
robust AF-correlations 
``across-the-hole'' (AH) (Fig.1a). 
Similar correlations were noticed
 in the standard $\rm t-J$ model on
ladders\cite{white1} and using small clusters
with NNN-hoppings\cite{tohyama2}. 
However, 
the physical origin of the AH-correlations,
the full spin arrangement around the hole, and specially its
consequences,
remain to be identified.
In particular, due to the AH-structure here it is argued that
the vicinity of the hole carries
 a small spin,
correlated with a tiny quasiparticle (q.p.) weight Z in the one-particle
spectral function, particularly at ${\bf k} = (\pi,0)-(0,\pi)$. This is suggestive of
spin-charge separation in two-dimensions,
at least at short distances, a phenomenon searched for
since the early proposals of $\rm high$-$\rm T_c$ theories.
The present results suggest that this effect may be  relevant at the 
couplings currently believed to be realistic for the cuprates.

 Our analysis starts with the 
observation that the one-hole ground-state q.p. weight Z
in the $\rm t-J$ model vanishes only
in the limit when J/t$\rightarrow$0
since the size of the spin-1/2 spin-polaron around the hole,
regulated by string excitations,
diverges as $\rm J \rightarrow 0$\cite{review}. 
However, even at J/t=0 a state with vanishing Z is not obtained
since in this limit the  one-hole 
ground-state becomes ferromagnetic (FM) due to the Nagaoka mechanism.
However, additional hole-hopping terms
may stabilize
a more exotic ground-state since the extra hole-mobility
can scramble severely 
the spin background near the hole, avoiding the localizing tendencies 
 of the string excitations.

\begin{figure}[htbp]
\centerline{\psfig{figure=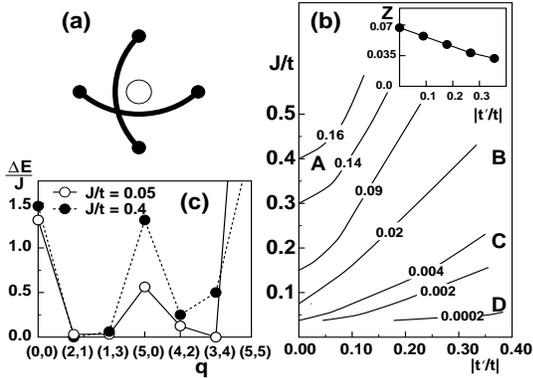,width=7.0cm,height=5.0cm,angle=-90}}
\vspace{0.3cm}
\caption{(a) Schematic representation of the 
AF-bonds across the hole (solid lines)
of the extended $\rm t-J$ 
model. Shown are five sites, with the hole at the
center; (b) Lines of constant Z corresponding to momentum
$(\pi,0)$, obtained using a 20-sites cluster with one-hole, 
a grid of 16
points in parameter space, and smooth interpolations among them,
providing sufficient accuracy for our qualitative discussion.
$\rm t'/t''$ is fixed to -1.4.
Points A, B, C and D
are mentioned in the text (B corresponds to the currently accepted
coupling for the cuprates).
In the inset, Z for an
18-site cluster and
${\bf k} = (\pi/3,\pi/3)$ is shown vs. $\rm |t'/t|$, at $\rm J/t=0.125$; 
(c)
Energy of a hole on a 20-sites cluster for the available momenta, 
relative to the one-hole ground-state energy. 
$\rm t'/t=-0.35, t''/t=+0.25$ and J/t's are
indicated. The momenta are in units of $\pi/5$.}
\label{fig1}
\end{figure}

To analyze such a possibility,  Z
has been calculated in the extended $\rm t-J$ model. 
Results are shown in Fig.1b, using 
Exact Diagonalization (ED) techniques
on 18- and 20-sites clusters\cite{review}. 
The ratio $\rm t'/t''$ was fixed to $-1.4$ in most of our study, 
as suggested by fits of
ARPES dispersions for $\rm Sr_2 Cu O_2 Cl_2$\cite{eder}. From Fig.1b
it is clear that adding amplitudes $\rm t'/t < 0$, $\rm
t''/t > 0$ drastically
reduces Z at ${\bf k } = (\pi,0)$, making it
virtually negligible 
in large regions of parameter space. 
The inset of Fig.1b shows that similar results are obtained 
close to ${\bf k} = (\pi/2,\pi/2)$, although the reduction of Z with
increasing NNN-hoppings is much stronger at $(\pi,0)$.
The shape of the constant-Z lines seems mainly regulated
by the strength of $\rm t',t''$ relative
to J, reasonable since $\rm t$ is 
renormalized to J by AF-fluctuations\cite{review},
while $\rm t',t''$ are 
only partially renormalized since they connect
same sublattice sites. Note that
the region of very small  $\rm J/t$ of the $\rm t-J$ model,
hidden by the FM instability,
is substantially expanded
by the addition of NNN-hoppings.

The remarkable isotropy around $(\pi/2,\pi/2)$
found in ARPES experiments\cite{wells,ronning} 
for $\rm Sr_2 Cu O_2 Cl_2$ 
is reproduced in the extended $\rm t-J$ model\cite{nazarenko,bel,eder,kim,refer},
and Fig.1c for a 20-sites lattice confirms that the four available
momenta near $(\pi/2,\pi/2)$ indeed have lower energy than
$(\pi,0)$, once the proper set of NNN-hoppings\cite{eder} is used.
However,
Fig.1c also shows that 
similar results are found
at smaller couplings J/t=0.05, keeping the NNN-hoppings the 
same\cite{vojta}. This suggests
that a quasi-isotropic dispersion
around $(\pi/2,\pi/2)$ may exist in the broad parameter region
with small Z. In this respect, points B, C  and
D of Fig.1b share similar properties, different from the
traditionally studied regime of point A. This will be a recurrent
conclusion of the results analyzed below.

\begin{figure}[htbp]
\centerline{\psfig{figure=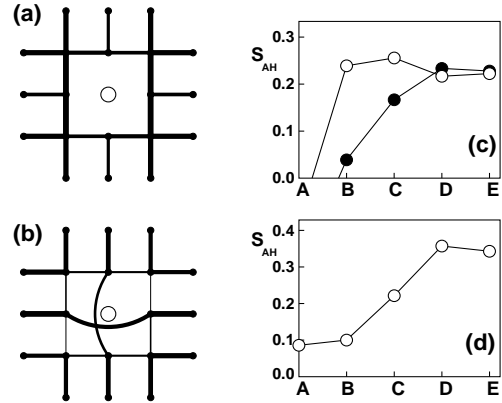,width=6.5cm,height=5.3cm,angle=-90}}
\vspace{0.3cm}
\caption{Spin-correlations in the mobile hole
reference frame
obtained using a 20-sites cluster with one-hole, and 
${\bf k} = (\pi,0)$. (a) corresponds to $\rm J/t=0.4$ 
and $\rm t'=t''=0$, while (b) is for 
$\rm J/t=0.2$, $\rm t'/t=-0.35$, $\rm t''/t=+0.25$. The width of the
lines is proportional to the strength of the AF-bonds. Only
AF-correlations at distances $a$ and $2a$ are shown; (c) Strength of the
AF-correlation across-the-hole $\rm S_{AH}$ 
for the ground-state of one-hole on a
 20-sites cluster  with ${\bf k} = (\pi,0)$. A positive (negative) result
represents an AF (FM) correlation.
Shown are data in the
x-direction (open) and y-direction (full). The points $\rm A,B,C,D$ 
(Fig.1b) and $\rm E$
correspond to $\rm (J/t=0.4,t'/t=0.0,t''/t=0.0)$, 
$\rm (0.4,-0.35,+0.25)$, $\rm (0.2,
-0.35, +0.25)$, $\rm (0.05, -0.35, +0.25)$, and $\rm (0.05,$ $\rm -1,+1)$,
respectively; (d) Same as (c) but on a 18-sites cluster with ${\bf k}
= (\pi/3,\pi/3)$. Here x- and y-directions are equivalent.}
\label{fig2}
\end{figure}

To understand the drastic reduction of Z 
after adding  NNN-hoppings, consider in Figs.2a-b the 
spin correlations in the
hole reference
frame. Shown are results 
for a 20-sites cluster exactly solved
and momentum $(\pi,0)$, of importance
for the cuprates.
Results are presented for points A and C of Fig.1b. 
Similar results have been obtained using the 
ORBA technique\cite{orba} 
on clusters of size $4 \times 8$ with $1.5 \times 10^6$ states,
and also
with ED on 16-sites clusters, suggesting
that finite-size effects are small. Fig.2a shows 
 that in the absence of NNN-hoppings
the spin correlations present simple AF tendencies\cite{bonca}. 
The hole carries a spin cloud
and Z is finite\cite{review}. 
However, including extra hoppings
the spin correlations are qualitatively different (Fig.2b). Note the 
presence of  AF-correlations
across the hole, in both directions. 
The hole motion induces a coupling between
sites of the same sublattice.
It is important to note that 
the across-the-hole AF-bond is not isolated but it is
supplemented in both directions by other
robust AF-bonds resembling dynamically generated 
 1D Heisenberg segments  along each axis (Fig.2b), 
individually weakly coupled to the rest of the spins.
Figs.2c,d show the strength of the 
AH-bond at several points in parameter space.
Comparing with Fig.1b, it is apparent that the
stronger
those correlations are, the weaker Z is.

The robust AF-correlations among pairs of spins near the hole
suggest that the total spin-1/2 
of the one-hole problem, defined on an even number of sites
 cluster, may not be located near the hole. 
This can be analyzed by calculating
the local spin $\rm \langle S^z_i \rangle $ at site $\rm  i$ 
in the mobile hole reference frame, with the overall constraint
$\rm \sum_i \langle S^z_i \rangle = 1/2 $. At $\rm J/t$=$0.4$, $\rm
t'$=$\rm t''$=$0$,
Fig.3a shows $\rm \langle S^z_i \rangle $ around the mobile hole for
${\bf k} = (\pi,0)$. Here the spin distribution
is nontrivial and some results
are even negative along the direction of movement of the hole.
However, the large spin next to the hole along the y-axis suggests that
spin and charge are bounded.
On the other hand, at $\rm J/t=0.1$, $\rm t'=-t''=-t$ (Fig.3b), 
here shown as an extreme but illustrative example,
$\rm \langle S^z_i\rangle$ is considerably more spread than in Fig.3a.
A robust AH-spin arrangement 
near the hole appears correlated with 
local spin-charge separation tendencies, 
which is intuitively reasonable. 

\begin{figure}[htbp]
\centerline{\psfig{figure=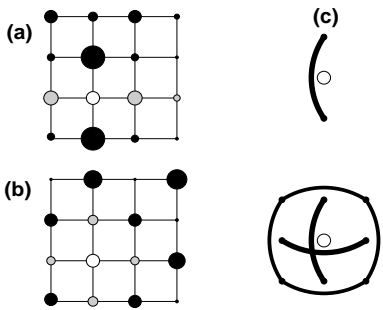,width=5.0cm,height=4.0cm,angle=-90}}
\vspace{0.3cm}
\caption{(a)  $\rm \langle S^z_i \rangle$ 
around a mobile hole (open circle) on a $4\times 4$ cluster
with one hole, total spin z-projection +1/2, $\rm J/t=0.4$, $\rm
t'=t''=0$, and ${\bf k} = (\pi,0)$. 
The area of the circles is proportional to $\rm \langle S^z_i \rangle$. 
The gray circles denote a negative $\rm \langle S^z_i \rangle$;
(b) Same as (a) but for 
$\rm J/t=0.1$, $\rm t'=-t''=-t$; 
(c) Schematic representation of the 3- and
9-site clusters mentioned in the text, the latter at $\rm J/t=0.1$, $\rm
t'/t=-0.35$ and $\rm t''/t=0.25$. The hole is at the center and the
solid lines represent the ground-state dominant AF-bonds, which are
 stable in a broad region of parameter space.}
\label{fig3}
\end{figure} 

To gain insight into the AH-state,
consider first just
3 sites in an open chain, 
containing one hole, one spin-up and one down (6 states). 
Since J is not crucial 
in this analysis simply use J=0, 
and allow for the hole to move at distance $a$ ($2a$) with
amplitude t ($\rm t''$). 
Two eigenstates are 
in competition for the ground-state. 
One has FM-correlations (Nagaoka state)
and the lowest energy for $\rm t'' < 0$. 
However, if
$\rm t'' > 0$ 
a spin-singlet state with energy 
$\rm E= {{1}\over{2}}(-t'' - \sqrt{(t'')^2 + 8 t^2 } )$
is stabilized (Fig.3c).
The spin-singlet nature of this state leads in a natural way to AH-bonds.
Then, the extra $\rm t'<0$ and $\rm t''>0$  hoppings
have the important role of favoring 
the singlet state over the
competing FM-state\cite{comm1,comm3}.

Further semi-quantitative insight can be gained from the analysis of a
9-site cluster with open boundary conditions
(Fig.3c). Here there are 630 states in the
zero total-$\rm S^z$ subspace, and characteristics already 
 similar to those found in the
numerical analysis of Figs.2a-b were observed in its solution.
The ground state of this small cluster is 
dominated by a hole at the center, in a large region of parameter 
space. 
The 
similarities of the ground-state correlations on  9-sites and on larger clusters
suggest, once again, that the effects discussed here are mainly local in
space and they can be observed in simple toy models.

Qualitative understanding of the AH-state formation can be obtained
 from the 1D-Hubbard
model at $\rm U/t=\infty$\cite{shiba}. In this limit holes and spins
fully decouple and the wave function in the spin sector corresponds to
that of a 1D-Heisenberg chain involving all spins, 
as if the holes were absent from the problem. 
In other words, a state such as 
$| ... + - + - 0 + - + 0 - + ...\rangle$, with AF-bonds across the
holes, has an important
weight in the ground state. 
Explicit calculations performed 
as part of this effort have indeed shown that the 1D $\rm
t-J-t'-t''$ model presents AH-bonds quite similar to those in Fig.2b. 
However, if a similar procedure is attempted in 2D,
i.e. the effective removal of the hole from the system by
linking the two vertical and horizontal neighboring spins with a
Heisenberg 
coupling J, frustration cannot be avoided. Then, a straightforward
generalization to 2D of the 
1D results is expected to fail. However, the numerical
results shown here lead us to suggest that a compromise can be found
between the tendencies
to produce a charge-spin decoupling and the 
frustration that prevents it.  This can be achieved by creating
1D-like spin arrangements in both directions, as shown in Fig.2b.
Each of these dynamically generated ``chains'' individually 
resemble the results
found in the 1D $\rm U/t = \infty$ Hubbard model\cite{shiba}.
By this procedure, 
spin and charge can locally decouple in 2D, as suggested by the
computational results.

The spin arrangement around the 2D hole at $\rm {\bf k} = (\pi,0)$
reported here is different from previous
alternatives discussed in the literature. 
Although averaging spin
correlations around the hole would only indicate
a reduction of antiferromagnetism
in favor of a ``spin-liquid'' state, the spin arrangement
in this region has more structure than a mere bubble of weakly
interacting spins would have.
In addition, while naively it may appear
that the AH-bond could be described as a spin-singlet
formation 
in a short-range resonant-valence-bond (RVB)
state, Fig.2b shows that 
it is better to represent it as an AF-bond part of a 1D
Heisenberg segment, an unexpected result in a 2D system. Note also that
moving holes using 
NNN-hoppings in a RVB-state 
would tend to leave the ``length'' of the short 
spin-singlets at just $a$, an
effect opposite to what was reported 
here, namely NNN-hoppings were found to enhance the AH-bonds.
Then, the AH-state 
represents a new paradigm for the visualization of the spin distortion
around a hole in an AF background.

\begin{figure}[htbp]
\centerline{\psfig{figure=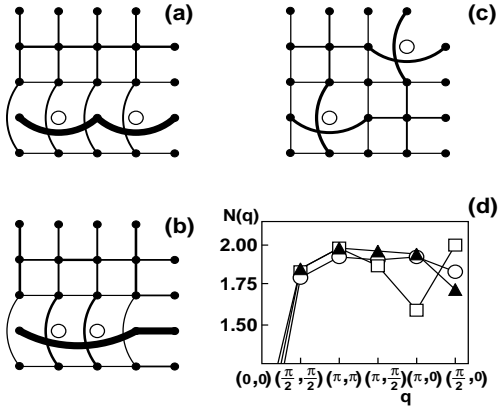,width=6.5cm,height=5.3cm,angle=-90}}
\vspace{0.3cm}
\caption{(a) AF-correlations corresponding to the $4 \times 4$ 
two-hole ground-state with PBC and
J/t=0.4, $\rm t'/t=-0.35$, $\rm t''/t=+0.25$, for the case when the
mobile holes (open circles) are
projected to be at distance $2a$ along the x-axis. The 
bond width is proportional to the strength of the
AF-correlation. The horizontal AH-bond strength is 
close to a perfect singlet, but this is due to the boundary conditions on the
16-sites cluster. In the bulk, its strength is expected to be reduced
roughly by a factor 2; (b) Same as (a) but with the holes
at distance $a$; (c) Same as (a) but with the holes
at a larger distance; (d) $\rm N(q)$ for a two-holes $4 \times 4$ cluster.
Squares, circles and triangles are results for $\rm (J/t=0.4$,$\rm
t'/t=0$,$\rm t''/t=0)$, $(0.4,-0.35,0.25)$, and $(0.1,-0.35,0.25)$, respectively.}
\label{fig4}
\end{figure}

In Figs.4a-c results are shown for two holes at 
point B of Fig.1b, the 
realistic couplings for the cuprates.
Fig.4a contains the spin
arrangement for two holes on a $4 \times 4$ cluster with PBC and 
at distance $2a$, obtained from the zero-momentum
ground-state using a suitable projector operator. With one hole at the
origin, the most likely position for the second hole is precisely at
distance $2a$, unlike for $\rm t',t''=0$ which is dominated by distance $a$.
The row where the holes are located is clearly different from the rest,
resembling a $4
\times 1$ cluster with two holes. This suggests that holes tend to move
in a 1D-path with density 0.5. 
These
1D-paths resemble metallic stripes, conceptually different from the
insulating stripes with one-hole per site found 
near the large J/t phase separated region of the standard 
$\rm t-J$ model when $\rm 1/r$-Coulomb interactions are added\cite{low}.
Fig.4b is a natural consequence of Fig.4a, when
holes move using the hopping t. 
An interesting property of Fig.4a-b is 
the arrangement of the spins not on the $4
\times 1$ stripe. They
form robust AF-bonds across the path of the two holes, as
observed experimentally\cite{tranquada}
and in ladder studies\cite{white1}.
This effect is not present
for holes with low-mobility near phase
separation, but in the context discussed here it
is natural since it evolves from
the AH-bonds of individual holes. 
Fig.4c corresponds to holes not along a
1D-path, carrying the surrounding 
environments of isolated holes. Fig.4d contains the
Fourier-transform of the charge correlations $\rm N(q)$, 
enhanced at 
$(\pi,0)$ due to configurations Figs.4a,b. Calculations for four holes
on 16-sites
show patterns similar to those in Fig.4. In particular, the
configuration with holes forming a square of side $2a$ has a substantial
weight in the ground-state and contains two of the
1D-like paths of Fig.4a, running in each direction.

It is
important to remark that for the standard $\rm t-J$ model with
J/t between 0.1 and 0.5 and $\rm t'=t''=0$
the correlations across-the-stripe are
actually FM on the 16-sites cluster, although
the spin correlation among the two spins of the two-holes 
$4\times 1$ row is still AF and robust. The 
NNN-hoppings are needed to stabilize a structure with
characteristics similar to those observed in experiments\cite{tranquada}.
Actually, results  quantitatively similar to
Figs.4a-c were obtained also at point C of Fig.1b, and even in an 
extreme case such as $\rm J/t=0.1$, 
$\rm t'=-t$, $\rm t''=t$.

Summarizing, here it has been reported that a mobile hole in the
extended $\rm t-J$ model generates a complex spin arrangement in its
vicinity containing AF-bonds across the hole. This is correlated with a
small Z and indications of spin-charge separation at $\rm {\bf k}= (\pi,0)$. 
Further work will clarify the relevance of the one-hole AH-bonds at
finite hole density. However, 
studies of two and four holes on small clusters have already
provided a glimpse of a possible new mechanism for metallic stripe
formation, which does not rely on phase separation tendencies and it is 
operative at small J/t.

The authors thank M. Vojta and T. Tohyama for many useful discussions.
NSF support (DMR-9814350 and NHMFL
In-House DMR-9527035) is acknowledged.

\medskip

\vfil


\begin{references}
\bibitem{comm1} This can be easily confirmed 
by an analysis of the Nagaoka theorem
when the NNN-couplings 
are introduced.
\bibitem{comm3} 
Similar results can be obtained on a 5-site
cluster (Fig.1a).
\end{references}
\end{document}